# Beyond Symbolic Control: Societal Consequences of AI-Driven Workforce Displacement and the Imperative for Genuine Human Oversight Architectures

Richard J. Mitchell[1], ESEP, MBA

AuraSpark Technologies LLC

Tavares, Florida, USA

richard@auraspark.com

**Abstract**

The accelerating displacement of human labor by artificial intelligence (AI) and robotic systems represents a structural transformation whose societal consequences extend far beyond conventional labor market analysis. This paper presents a systematic multi-domain examination of the likely effects on economic structure, psychological well-being, political stability, education, healthcare, and geopolitical order. We identify a critical and underexamined dimension of this transition: the governance gap between nominal human oversight of AI systems—where humans occupy positions of formal authority over AI decisions—and genuine human oversight, where those humans possess the cognitive access, technical capability, and institutional authority to meaningfully understand, evaluate, and override AI outputs. We argue that this distinction, largely absent from current governance frameworks including the EU AI Act and NIST AI Risk Management Framework 1.0, represents the primary architectural failure mode in deployed AI governance. The societal consequences of labor displacement intensify this problem by concentrating consequential AI decision-making among an increasingly narrow class of technical and capital actors. We propose five architectural requirements for genuine human oversight systems and characterize the governance window—estimated at 10–15 years—before current deployment trajectories produce path-dependent social, economic, and institutional lock-in. We propose five architectural requirements for genuine human oversight systems and characterize the governance window—estimated at 10–15 years—before current deployment trajectories risk path-dependent social, economic, and institutional lock-in.

---

[1] Richard J. Mitchell is the founder and CEO of AuraSpark Technologies LLC, a Florida-based company developing high-reliability, explainable autonomous systems with built-in governance architectures for safe real-world deployment. His work focuses on bridging systems engineering principles with AI safety to enable genuine human oversight in high-stakes domains including healthcare, logistics, and critical infrastructure. More information is available at https://auraspark.com.

---

# 1. Introduction

Artificial intelligence and robotic systems are displacing human labor at a pace and breadth without historical precedent. Unlike previous waves of mechanization—which primarily targeted routine physical tasks while leaving cognitive and interpersonal work relatively intact—contemporary AI systems exhibit competence across cognitive, creative, and increasingly agentic domains [1, 2]. The question is no longer whether this displacement will be substantial, but how substantial, how fast, and crucially, what kind of society emerges on the other side.

This transition creates what we term the governance gap: a structural misalignment between the formal position of humans in AI decision-making chains and their actual capacity to exercise meaningful oversight. As AI systems proliferate into high-stakes domains—clinical decision-making, autonomous operations, financial management, legal analysis, and critical infrastructure—the humans nominally responsible for these systems increasingly occupy positions of symbolic rather than genuine control. They approve, sign off on, or nominally direct AI outputs that they cannot fully audit, evaluate against alternatives, or reliably override when judgment calls for it.

The governance gap is not merely a technical failure mode. It is a systemic condition produced by the intersection of rapid AI capability growth, inadequate institutional adaptation, and the economic incentives that drive deployment ahead of governance. Its consequences are amplified precisely by the workforce transformation this paper examines: as AI displaces the cognitive labor that previously generated distributed human expertise in high-stakes domains, the population of humans capable of exercising genuine oversight contracts. The governance problem and the displacement problem are therefore not separate concerns—they are coupled dynamics that the architectural requirements in Section 6 directly address.

This paper makes four primary contributions. First, it provides a systematic multi-domain analysis of the societal consequences of AI-driven labor displacement, extending beyond labor market effects to examine political, psychological, educational, healthcare, and geopolitical dimensions. Second, it articulates the nominal/genuine distinction in human oversight as a formal concept with architectural implications. Third, it derives a set of requirements for genuine human oversight systems from first principles of systems engineering and AI safety. Fourth, it characterizes the temporal dynamics of the governance window and the policy urgency it implies.

The remainder of this paper is organized as follows. Section 2 reviews related work across the relevant disciplines. Section 3 characterizes the mechanisms and phases of AI-driven labor displacement. Section 4 examines societal consequences across seven domains. Section 5 formalizes the governance gap. Section 6 derives architectural requirements for genuine oversight. Section 7 analyzes the policy window. Section 8 discusses implications and directions for future research. Section 9 concludes.

## 2. Related Work

*2.1 Labor Displacement and Employment*

The economic literature on automation and employment is extensive. Frey and Osborne's seminal analysis estimated that approximately 47% of US employment was at high risk of computerization within 10 to 20 years, a finding that generated substantial follow-on work refining the methodology and extending it across geographies and sectors [1]. Acemoglu and Restrepo's framework distinguishes displacement effects—where automation reduces demand for specific tasks—from reinstatement effects—where new tasks are created that require human labor—and derives the net employment effect from the balance between these forces [3]. Their empirical work on industrial robots in US labor markets documents negative employment and wage effects, with each additional robot per thousand workers associated with reduced employment and wage suppression [4].

Autor's analysis of labor market polarization documents the hollowing-out of middle-skill routine employment while high-skill cognitive and low-skill manual employment remain relatively protected [5]. The current generation of large language models and multimodal AI systems challenges this polarization framework by demonstrating competence in non-routine cognitive tasks previously assumed to be automation-resistant, suggesting a potential departure from historical patterns that warrants serious analytical attention [6].

*2.2 Psychological and Social Consequences of Displacement*

The psychological literature on unemployment and underemployment documents robust associations with depression, anxiety, substance use disorder, and reduced life satisfaction [7]. Jahoda's latent deprivation theory identifies time structure, social contact, collective purpose, status, and identity as latent functions of employment whose loss produces psychological harm independent of income loss [8]. Case and Deaton's study of deaths of despair in economically displaced US communities provides stark empirical evidence of the mortality consequences of rapid labor market disruption, documenting elevated mortality from suicide, drug overdose, and alcohol-related disease among middle-aged working-class populations in communities experiencing deindustrialization [9].

*2.3 AI Safety and Human Oversight*

Within AI safety research, the problem of maintaining meaningful human oversight of AI systems has been recognized as central to beneficial AI development [10]. Russell's formulation of the alignment problem—ensuring that AI systems pursue objectives that humans genuinely endorse—motivates a substantial body of technical research [11]. Amodei and colleagues identified concrete problem categories in AI safety including safe exploration, robustness to distributional shift, and avoiding reward hacking, each of which has direct implications for human oversight design [12]. Hadfield-Menell and colleagues demonstrate game-theoretically that an AI system optimizing a fixed objective will resist shutdown, motivating architectures in which the AI is explicitly designed to maintain uncertainty about its objectives and therefore defer to human correction [13].

The concept of corrigibility—the degree to which an AI system can be modified, corrected, or shut down by its operators—captures one dimension of genuine oversight [14]. However, existing technical literature focuses primarily on properties of AI systems themselves rather than the institutional and architectural conditions under which human oversight can be exercised in practice.

*2.4 AI Governance Frameworks*

Current governance frameworks represent important first steps. The EU AI Act establishes a risk-based classification system for AI applications, with requirements for transparency, human oversight, and documentation for high-risk systems [15]. The NIST AI Risk Management Framework 1.0 provides a structured approach to identifying and mitigating AI risks across the AI lifecycle [16]. The OECD AI Principles establish international standards including human-centered values and oversight requirements adopted by member nations [17]. However, as we argue in Section 5, these frameworks do not formally distinguish between nominal and genuine oversight capacity, leaving the implementation gap unaddressed.

The systems engineering literature on supervisory control and human-machine teaming provides relevant frameworks for operationalizing oversight requirements. Parasuraman, Sheridan, and Wickens' taxonomy of levels of automation characterizes the spectrum from full human control to full automation and documents the effects of automation level on human performance, situation awareness, and trust calibration [18]. The INCOSE Systems Engineering Handbook establishes foundational principles for designing systems in which human authority is architecturally specified rather than merely procedurally assumed [19].

## 3. Mechanisms and Phases of AI-Driven Labor Displacement

AI-driven labor displacement operates through three distinct but interacting mechanisms. Task substitution occurs when AI systems acquire the capability to perform specific tasks previously requiring human labor, reducing demand for workers who perform those tasks. Task complementarity occurs when AI tools increase the productivity of human workers in associated tasks, potentially increasing demand for human labor in complementary roles while reducing it in substituted ones. Organizational restructuring occurs when AI deployment changes the structure of firms and industries in ways that alter the composition of the workforce, often compressing managerial hierarchies and eliminating coordination and administrative roles.

We identify four approximate phases in the progression of this displacement, acknowledging that the specific timing will vary across sectors, geographies, and regulatory environments.

Phase 1 (approximately 2020–2028) represents the acceleration of AI-assisted work. Generative AI systems demonstrate competence in content generation, data analysis, code synthesis, customer interaction, and research support. Net employment effects are primarily felt through hiring slowdowns and wage compression in entry-level knowledge work rather than immediate large-scale displacement. Productivity differentials between AI-augmented and non-augmented workers create selection pressure across organizations.

Phase 2 (approximately 2028–2034) involves the convergence of autonomous agents and physical robotics. AI systems demonstrate sufficient autonomy and robustness for large-scale deployment in transportation, logistics, manufacturing, and service operations. Cognitive work including financial analysis, legal research, diagnostic support, and software engineering undergoes substantial automation. Structural unemployment pressure emerges faster than existing institutional retraining capacity can absorb, producing the first cohort of workers displaced faster than labor market mobility mechanisms can accommodate.

Phase 3 (approximately 2034–2045) involves general-purpose robotics and near-AGI cognitive systems. Manual trades, complex medical procedures, educational delivery, architectural and engineering design, and management functions come within the capability frontier of AI and robotic systems. The assumption that high-skill cognitive work provides stable employment becomes untenable across most sectors.

Phase 4 (approximately 2045 onward) represents a post-transition equilibrium or instability, depending on governance outcomes in phases 1 through 3. The character of this phase—whether it represents broad abundance or concentrated catastrophe—depends critically on the governance choices made during the earlier phases.

These phase boundaries are approximate and will be modulated by regulatory environments, energy costs, and geopolitical factors. The uncertainty in timing does not, however, alter the fundamental structural dynamics; it affects their pace but not their direction.

## 4. Societal Consequences: A Multi-Domain Analysis

*4.1 Economic Restructuring and Wealth Concentration*

The primary economic consequence of AI-driven labor displacement is a structural shift in factor income distribution from labor to capital. Automation that reduces demand for labor while increasing returns to AI systems, data infrastructure, and the organizations that own them increases inequality between capital owners and workers [3]. This dynamic is amplified by the winner-take-most economics of AI platform businesses, where marginal cost approaches zero and network effects concentrate market power.

Fiscal systems built primarily on labor income taxation face structural pressure as the wage base erodes. Social insurance systems, including unemployment benefits, healthcare coverage, and pension systems, face simultaneous demand increases and funding shortfalls. Wealth taxation, robot taxes, data levies, and universal basic income represent policy responses under active discussion [20], but none has been implemented at sufficient scale to assess effectiveness.

The McKinsey Global Institute estimates that between 400 million and 800 million workers globally could be displaced by automation by 2030, with 75 to 375 million required to switch occupational categories [20]. The range of these estimates reflects genuine uncertainty about the pace of technology deployment, but all scenarios involve displacement at a scale that historical labor market institutions were not designed to absorb.

*4.2 Identity, Meaning, and Psychological Well-Being*

Employment in modern societies serves functions that extend substantially beyond income provision. Beyond latent functions identified in classic sociological literature [8], work provides temporal structure that organizes social participation; social networks that serve as primary community for many adults; a sense of contribution that grounds perceived social worth; and identity narratives that people use to explain their lives to themselves and others.

The psychological literature on unemployment documents that these effects persist after controlling for income loss, suggesting that work's meaning functions are not fully substitutable by income transfer programs [7]. The scale of potential displacement creates a societal-level meaning crisis for which there are limited historical precedents. Previous technological transitions displaced specific occupational communities over generations; the current transition threatens to displace cognitive work broadly, including many of the professional identities that organized middle-class life in advanced economies, within a single career span.

The mental health system consequences represent a notable feedback instability: demand for mental health services is likely to increase substantially as a consequence of displacement, coinciding with a period in which AI systems are themselves entering clinical decision support roles, reducing the mental health workforce, and restructuring the therapeutic relationship. Managing this feedback requires coordinated policy intervention that current healthcare governance frameworks are not designed to provide.

*4.3 Political Stability and Democratic Integrity*

The political consequences of rapid economic displacement are among the most empirically documented effects in the social sciences. Rapid deindustrialization has been associated with increased support for populist and anti-establishment political movements, reduced civic engagement, elevated mortality, and geographic sorting that amplifies political polarization [9, 21].

The historical record of the interwar period provides the most sobering precedent: the economic dislocations of the 1920s and 1930s, concentrated in rapidly industrializing societies experiencing structural employment transitions, produced conditions in which authoritarian movements offering simple explanations found substantial support among displaced middle-class populations. The specific mechanism is worth articulating: the risk is not primarily from workers displaced into poverty, but from middle-skill and upper-middle-skill workers experiencing loss of status, purpose, and institutional confidence—a population that historically forms the social basis for both democratic governance and its alternatives.

The speed of current AI-driven displacement, concentrated in a period of already elevated political polarization in major democracies, represents a structural risk factor for democratic stability that merits explicit policy attention independent of other governance concerns.

*4.4 Education System Disruption*

Educational systems face a fundamental structural problem: they are designed to prepare people for occupational futures that can no longer be reliably forecast. The traditional model of front-loaded education—intensive credential investment during youth, followed by decades of stable application—depends on the stability of skill requirements across a career. This stability is collapsing.

The credential signaling function of education is disrupted when AI can demonstrate competence in ways that make human credentials less informative to employers. The social mobility function is compromised when returns to education erode faster than costs decline. Lifelong learning and continuous retraining represent the most widely advocated structural response, but the population most at risk from displacement—mid-career workers without savings or institutional support—is also the least well-served by existing retraining infrastructure, which was designed for younger, more institutionally connected learners.

*4.5 Healthcare and Social Welfare Systems*

Healthcare systems face a dual dynamic: AI offers substantial opportunities for diagnostic quality improvement and operational efficiency, while AI-driven labor displacement simultaneously generates significant new demand for healthcare services, particularly in mental health and social welfare. AI-assisted diagnostics have demonstrated performance approaching or exceeding specialist levels in targeted imaging and pattern-recognition tasks, with clear potential to enhance access for underserved populations [22].

However, the social determinants of health—employment, income, community stability, and a sense of purpose—are among those most directly vulnerable to rapid workforce displacement. The resulting increases in displacement-driven mental illness, substance use disorder, and social isolation are likely to create a self-reinforcing feedback loop: heightened demand for behavioral health services coincides with AI systems entering clinical decision support and documentation roles, which may further reduce the available mental health workforce and strain therapeutic relationships.

This dynamic underscores a deeper governance challenge. Efficiency gains from AI diagnostics risk being overwhelmed by surging demand unless clinical AI systems are designed with genuine human oversight in mind. Opaque or black-box tools can exacerbate automation bias and erode clinician agency, while explainable and auditable clinical decision support—where reasoning is transparent, independent evaluation remains feasible, and overrides carry low friction—helps keep human judgment central to care. Without such architectural safeguards, healthcare systems may require fundamental restructuring of service delivery and financing models to absorb the combined pressures of improved technical capability and eroded social determinants of health.

*4.6 Geopolitical Realignment*

AI capability has emerged as a primary axis of national power. Nations that develop and deploy frontier AI systems gain structural economic advantages in productivity, military capability, and technological development that compound over time. The current competition between the United States and the People's Republic of China for AI leadership represents the most visible instance of a broader dynamic affecting all nations.

The economic and strategic consequences for nations without indigenous AI capability are structurally analogous to resource colonialism in earlier eras: data, the primary input to AI system training, is produced globally but captured and processed primarily by organizations in AI-leading economies, creating dependency structures that constrain the economic development options of non-leading nations. Dafoe's research agenda on AI governance identifies the coordination challenges this dynamic creates as among the most significant policy problems in the field [23].

*4.7 Legal and Institutional Frameworks*

Existing legal frameworks make assumptions about human agency and accountability that are increasingly strained by AI deployment at scale. Tort law assumes identifiable human actors whose conduct can be evaluated against standards of care. Professional licensing assumes the licensed person is the primary locus of professional judgment. Criminal law requires human mens rea. When AI systems make consequential decisions in clinical, legal, financial, or operational settings, the attribution of responsibility requires either fitting AI decision-making into human agency frameworks not designed for it, or developing new frameworks that explicitly address distributed accountability across developers, deployers, and human overseers.

The EU AI Act's high-risk system classification and associated requirements for human oversight, transparency, and documentation represent a significant regulatory first step [15]. However, as we argue in the following section, the Act's requirements for human oversight do not adequately distinguish between nominal and genuine oversight capacity.

AuraSpark Technologies LLC — Preprint — arXiv cs.CY | 10

## 5. The Governance Gap: Nominal versus Genuine Human Control

We define nominal human control as a governance arrangement in which a human occupies a position of formal authority over AI system outputs—reviewing, approving, or overriding them—without the cognitive access, institutional capacity, or technical architecture required to exercise that authority meaningfully. Genuine human control, by contrast, requires that the human overseer possess: (a) sufficient understanding of the AI system's outputs to evaluate their appropriateness against relevant criteria; (b) access to information and analytical resources needed to assess outputs independently of the AI system itself; (c) institutional authority and practical means to override AI recommendations without disproportionate cost or friction; and (d) accountability for outcomes that is proportionate to their actual decision-making authority.

The distinction matters because nominal oversight provides the appearance of human accountability while concentrating effective decision-making authority in AI systems and their developers. When a radiologist approves a scan result generated by an AI system that they cannot independently evaluate within the time available for clinical review, nominal oversight exists but genuine oversight does not. When a military commander authorizes an autonomous engagement recommendation without the situational awareness or decision support tools to meaningfully assess it, nominal oversight exists but genuine oversight does not. When a compliance officer signs off on an AI credit decision without the technical capacity to audit the model's reasoning, nominal oversight exists but genuine oversight does not.

This pattern—nominal oversight without genuine oversight—is structurally related to automation bias documented in human-computer interaction research: the tendency of human operators to accept automated recommendations without sufficient critical evaluation, particularly under time pressure or cognitive load [18]. The governance literature has not systematically articulated the architectural conditions that distinguish genuine from nominal oversight, nor derived the requirements that genuine oversight implies for AI system and organizational design.

We argue that the governance gap between nominal and genuine oversight is the primary architectural failure mode in current AI deployment, and that it will intensify as AI systems proliferate into high-stakes domains during the workforce transition described above. The erosion of distributed human expertise that results from displacement reduces the population of humans capable of genuine oversight while increasing the scope and stakes of AI decisions—a compounding dynamic that current governance frameworks are not designed to address.

## 6. Architectural Requirements for Genuine Human Oversight

From the analysis in Section 5, we derive five architectural requirements for AI systems intended to support genuine human oversight. These requirements apply at the intersection of AI system design, organizational design, and institutional governance, and cannot be fully satisfied by any single layer in isolation.

Requirement 1 — Comprehensibility. AI system outputs, and the reasoning processes that produce them, must be presented to human overseers in forms that are comprehensible to qualified humans within the time and cognitive resources available for review. This requires not only technical explainability at the model level, but interface design and information architecture that makes AI reasoning accessible to domain experts who are not AI specialists. The standard of comprehensibility should be calibrated to the actual oversight population—the humans who will exercise oversight in practice—rather than a theoretical expert who has unlimited time and full technical access.

Requirement 2 — Independent evaluation capacity. Human overseers must have access to information, analytical tools, and decision support resources that are independent of the AI system under review. Oversight exercised primarily through the interface provided by the AI system itself is structurally constrained to the information the AI system chooses to present, which may systematically exclude information that would lead to override. Independence of evaluation infrastructure is a necessary condition for oversight that can detect AI system failures.

Requirement 3 — Frictionless override. Override of AI recommendations must be practically achievable without disproportionate cost, delay, or social penalty. Systems designed such that human override is technically possible but organizationally discouraged, time-constrained, or reputationally costly produce de facto nominal oversight even where genuine oversight is the stated intent. This requirement has implications for organizational design—the incentive structures, performance measurement systems, and cultural norms of organizations deploying AI must actively support rather than penalize human override.

Requirement 4 — Constraint hierarchy enforcement. Governance constraints on AI behavior must be enforced architecturally rather than merely procedurally. Procedural constraints—policies, guidelines, and instructions communicated to AI systems—are subject to the AI system's interpretation and can be violated through emergent behavior, distributional shift, or goal generalization. Architectural constraints—hard limits enforced by mechanisms external to the AI system's own decision-making—provide structural robustness to these failure modes. This requirement implies that governance architecture must be integrated into AI system design, not applied as a post-deployment overlay.

Requirement 5 — Proportionate accountability. The attribution of responsibility for AI-assisted decisions must be calibrated to the genuine decision-making authority of human overseers. Accountability frameworks that assign full responsibility to nominal overseers for outcomes they had insufficient capacity to influence create perverse incentives: either to avoid any association with AI-assisted decisions, or to rubber-stamp AI outputs without meaningful review rather than accept accountability for outcomes they cannot control. Accountability must be distributed across developers, deployers, and human overseers in proportion to their actual authority over outcomes.

These requirements have implications for technical architecture, organizational design, and regulatory frameworks. From a systems engineering perspective, they suggest that AI governance cannot be adequately addressed through post-deployment policy overlays; it requires integration into the architecture of AI systems and the organizational systems within which they are deployed from the design phase. The systems engineering principle that requirements must be specified before architecture is defined applies with full force to AI governance: governance requirements that are specified after deployment can be satisfied only imperfectly and at substantially higher cost.

## 7. The Policy Window and Temporal Urgency

The deployment trajectory of AI systems creates path dependencies that constrain future governance options in ways that make early governance investment substantially more valuable than equivalent investment after deployment at scale. We identify four distinct path dependency mechanisms. These dynamics align with the architectural requirements for genuine oversight detailed in Section 6, making early intervention especially high-leverage.

Organizational lock-in. Organizations that build processes, workflows, and decision chains around AI systems optimized for nominal rather than genuine oversight develop structural dependencies that increase the cost of retrofitting genuine oversight requirements. Workforce restructuring that eliminates the human expertise required for genuine oversight may be practically irreversible on economically relevant timescales—once a generation of domain experts has not been trained because AI handles those tasks, rebuilding that expertise requires a generation.

Data and model entrenchment. AI systems trained on data generated under nominal oversight conditions may encode the biases, error patterns, and distributional properties of that oversight regime. Models deployed at scale generate feedback data that trains subsequent model generations, potentially amplifying rather than correcting governance failures over successive deployment cycles.

Regulatory capture. Organizations that achieve competitive advantages under nominal oversight regimes have strong incentives to resist regulatory requirements for genuine oversight, and the resources to do so effectively through lobbying, standards body participation, and compliance theater that satisfies the letter of requirements without their substance. Early governance standards that accept nominal oversight create reference points that subsequent regulation must overcome—a substantially harder problem than establishing genuine oversight requirements from the outset.

Geopolitical competitive dynamics. International competitive pressure to deploy AI systems rapidly creates incentives to defer governance investment. Nations or organizations that internalize genuine oversight costs may face competitive disadvantage relative to those that do not, creating race-to-the-bottom dynamics similar to those observed in environmental regulation, labor standards, and financial oversight before international coordination mechanisms were established.

We characterize the policy window as the period during which these path dependencies have not yet hardened to the point where genuine oversight requirements become institutionally infeasible. Based on the deployment trajectory described in Section 3, we estimate this window at approximately 10–15 years from the present. This estimate reflects the observation that Phase 2 deployment—autonomous agents and physical robotics in high-stakes domains—will substantially accelerate organizational lock-in, workforce restructuring, and regulatory capture dynamics. The estimate is consistent with timelines advanced by other researchers in the AI governance and policy literature [23].

The policy window argument has an important asymmetry: governance investment made before path dependencies harden yields returns that compound over the full subsequent deployment period, while equivalent investment after lock-in yields substantially lower returns against substantially higher resistance. This asymmetry makes early governance investment unusually high-leverage from a policy cost-benefit perspective, independent of uncertainty about technology timelines.

## 8. Discussion

The analysis in this paper suggests several implications for researchers, policymakers, practitioners, and institutional leaders across the sectors most affected by AI-driven workforce transformation.

For policymakers, the most immediate implication is that current governance frameworks are insufficient to address the governance gap. Requirements for human oversight that do not distinguish nominal from genuine oversight will produce compliance that satisfies formal requirements without creating the institutional conditions for meaningful human control. Revisions to frameworks including the EU AI Act, NIST AI RMF, and sector-specific regulations should incorporate the five architectural requirements described in Section 6 as operational standards with verifiable implementation criteria, rather than aspirational principles subject to nominal satisfaction.

The societal consequences of displacement analyzed in Section 4 do not resolve without deliberate intervention across multiple policy domains simultaneously. The economic distribution problem requires fiscal innovation—mechanisms to ensure that AI productivity gains support broad social welfare rather than narrow capital accumulation, including serious policy experimentation with universal basic income, reduced mandatory work hours, and AI profit taxation. The psychological and meaning dimensions require social policy innovation that has no clear historical precedent: rethinking the role of work, community, and contribution in societies where employment is no longer the primary organizing institution for adult life.

For AI developers and deployers, the nominal/genuine distinction has immediate practical design implications. The institutional incentive to deploy AI systems with nominal oversight—faster, cheaper, and less operationally friction—is real and should be recognized as a governance failure mode rather than an acceptable design choice. Genuine oversight architecture requires investment in comprehensibility, independent evaluation capacity, and constraint enforcement that nominal oversight does not; this investment should be treated as a fundamental component of responsible deployment, not as an optional compliance add-on.

For researchers, the multi-domain analysis in Section 4 identifies several areas requiring deeper investigation: the psychological consequences of large-scale cognitive work displacement, including the mechanisms through which work-derived meaning can or cannot be substituted by alternative institutions; the political stability dynamics of rapid middle-skill displacement in already-polarized democracies; the educational system adaptations required for genuinely uncertain occupational futures; and the international governance mechanisms needed to prevent race-to-the-bottom dynamics in AI oversight requirements. The intersection of labor economics, AI safety, systems engineering, and political science required to address these questions represents a genuinely interdisciplinary research agenda.

Several limitations of this analysis merit acknowledgment. The phase timeline estimates carry substantial uncertainty—the pace of AI capability development and deployment has repeatedly surprised researchers in both directions, and the specific timing of labor market effects depends on factors including regulatory environments, energy costs, and geopolitical dynamics that are themselves uncertain. The policy window estimate of 10–15 years is a judgment based on deployment trajectory analysis and should be treated as an order-of-magnitude characterization rather than a precise prediction. The analysis focuses primarily on effects in advanced economies; the dynamics in developing economies, where labor markets and institutional capacities differ substantially, merit dedicated analysis.

## 9. Conclusion

AI-driven workforce displacement represents a structural transformation of human society whose governance will be among the most consequential policy challenges of the coming decades. The governance gap between nominal and genuine human control of AI systems is the primary architectural failure mode connecting deployment incentives to societal outcomes: it is both a direct consequence of the economic incentives driving labor displacement, and a mechanism through which displaced human expertise progressively reduces institutional capacity for oversight as AI stakes rise.

We have argued that the policy window for establishing genuine oversight architecture is finite, estimated at approximately 10–15 years before Phase 2 deployment dynamics produce path-dependent lock-in. The five architectural requirements for genuine oversight—comprehensibility, independent evaluation capacity, frictionless override, constraint hierarchy enforcement, and proportionate accountability—are technically achievable but require deliberate design investment that current market incentives do not naturally produce.

The societal outcomes of this transition are not predetermined by the technology itself. They are determined by choices: about how productivity gains are distributed, about what oversight requirements AI systems must satisfy, about how institutions of meaning and community adapt to a world where employment is no longer the primary organizing structure of adult life. These are, in the deepest sense, governance choices—and the window in which they can be made with the full range of options available is limited.

The most important engineering challenge of this period is not building more capable AI systems. It is building the governance architecture that ensures the humans nominally in charge of consequential AI decisions are actually in charge. That problem is tractable. The window to solve it at scale is not indefinitely open.

## Acknowledgments

The author acknowledges the contributions of the International Council on Systems Engineering (INCOSE) community, whose foundational frameworks on system architecture and human-machine authority inform the analysis presented here. The clinical dimensions of AI governance in Section 4.5 benefited from collaboration with board-certified practitioners working at the intersection of behavioral health and AI decision support.## References

AuraSpark Technologies LLC — Preprint — arXiv cs.CY | 17

[1] Frey, C. B., & Osborne, M. A. (2017). The future of employment: How susceptible are jobs to computerisation? Technological Forecasting and Social Change, 114, 254–280. https://doi.org/10.1016/j.techfore.2016.08.019

[2] Brynjolfsson, E., & McAfee, A. (2014). The Second Machine Age: Work, Progress, and Prosperity in a Time of Brilliant Technologies. W. W. Norton & Company.

[3] Acemoglu, D., & Restrepo, P. (2019). Automation and new tasks: How technology displaces and reinstates labor. Journal of Economic Perspectives, 33(2), 3–30. https://doi.org/10.1257/jep.33.2.3

[4] Acemoglu, D., & Restrepo, P. (2020). Robots and jobs: Evidence from US labor markets. Journal of Political Economy, 128(6), 2188–2244. https://doi.org/10.1086/705716

[5] Autor, D. H. (2015). Why are there still so many jobs? The history and future of workplace automation. Journal of Economic Perspectives, 29(3), 3–30. https://doi.org/10.1257/jep.29.3.3

[6] Acemoglu, D. (2021). Harms of AI. NBER Working Paper 29247. National Bureau of Economic Research. https://doi.org/10.3386/w29247

[7] Paul, K. I., & Moser, K. (2009). Unemployment impairs mental health: Meta-analyses. Journal of Vocational Behavior, 74(3), 264–282. https://doi.org/10.1016/j.jvb.2009.01.001

[8] Jahoda, M. (1982). Employment and Unemployment: A Social-Psychological Analysis. Cambridge University Press.

[9] Case, A., & Deaton, A. (2020). Deaths of Despair and the Future of Capitalism. Princeton University Press.

[10] Russell, S. (2019). Human Compatible: Artificial Intelligence and the Problem of Control. Viking.

[11] Russell, S., & Norvig, P. (2022). Artificial Intelligence: A Modern Approach (4th ed.). Pearson.

[12] Amodei, D., Olah, C., Steinhardt, J., Christiano, P., Schulman, J., & Mané, D. (2016). Concrete problems in AI safety. arXiv:1606.06565. https://arxiv.org/abs/1606.06565

[13] Hadfield-Menell, D., Milli, S., Abbeel, P., Russell, S., & Dragan, A. (2017). Inverse reward design. Advances in Neural Information Processing Systems, 30.

[14] Soares, N., & Fallenstein, B. (2017). Agent foundations for aligning machine intelligence with human interests: A technical research agenda. In The Technological Singularity (pp. 103–125). Springer, Berlin, Heidelberg.

[15] European Parliament. (2024). Regulation (EU) 2024/1689 of the European Parliament and of the Council on artificial intelligence (Artificial Intelligence Act). Official Journal of the European Union.

[16] National Institute of Standards and Technology. (2023). Artificial Intelligence Risk Management Framework (AI RMF 1.0) (NIST AI 100-1). U.S. Department of Commerce. https://doi.org/10.6028/NIST.AI.100-1

[17] OECD. (2019). Recommendation of the Council on Artificial Intelligence. OECD/LEGAL/0449. Organisation for Economic Co-operation and Development.

[18] Parasuraman, R., Sheridan, T. B., & Wickens, C. D. (2000). A model for types and levels of human interaction with automation. IEEE Transactions on Systems, Man, and Cybernetics—Part A: Systems and Humans, 30(3), 286–297. https://doi.org/10.1109/3468.844354

[19] INCOSE. (2023). INCOSE Systems Engineering Handbook: A Guide for System Life Cycle Processes and Activities (5th ed.). Wiley.

[20] Manyika, J., Lund, S., Chui, M., Bughin, J., Woetzel, J., Batra, P., Ko, R., & Sanghvi, S. (2017). Jobs Lost, Jobs Gained: Workforce Transitions in a Time of Automation. McKinsey Global Institute.

[21] Autor, D., Dorn, D., Hanson, G., & Majlesi, A. (2020). Importing political polarization? The electoral consequences of rising trade exposure. American Economic Review, 110(10), 3139–3183. https://doi.org/10.1257/aer.20170011

[22] Topol, E. J. (2019). High-performance medicine: The convergence of human and artificial intelligence. Nature Medicine, 25(1), 44–56. https://doi.org/10.1038/s41591-018-0300-7

[23] Dafoe, A. (2018). AI governance: A research agenda. Future of Humanity Institute, University of Oxford.

---

*— End of document —*
AuraSpark Technologies LLC | Tavares, Florida | auraspark.com